\documentclass[%
 reprint,
superscriptaddress,
nofootinbib,
 amsmath,amssymb,
 aps,
prl,
]{revtex4-2}

\usepackage{graphicx}% Include figure files
\usepackage{subcaption}
\usepackage{dcolumn}% Align table columns on decimal point
\usepackage{bm}% bold math
\usepackage{siunitx}% non italicized mu for micro
\usepackage{soul}
\begin{document}

\title{Photon bunching in cathodoluminescence induced by indirect electron excitation}

\author{Vasudevan Iyer}
\affiliation
{Center for Nanophase Materials Sciences, Oak Ridge National Laboratory, Oak Ridge, TN, 37831, USA}
\author{Kevin Roccapriore}
\affiliation
{Center for Nanophase Materials Sciences, Oak Ridge National Laboratory, Oak Ridge, TN, 37831, USA}
\author{Jacob Ng}
\affiliation
{Technical University of Denmark, 2800 Kongens Lyngby, Denmark}
\author{Bernadeta Srijanto}
\affiliation
{Center for Nanophase Materials Sciences, Oak Ridge National Laboratory, Oak Ridge, TN, 37831, USA}
\author{David Lingerfelt}
\affiliation
{Center for Nanophase Materials Sciences, Oak Ridge National Laboratory, Oak Ridge, TN, 37831, USA}
\author{Benjamin Lawrie}
\affiliation
{Center for Nanophase Materials Sciences, Oak Ridge National Laboratory, Oak Ridge, TN, 37831, USA}
\affiliation
{Materials Science and Technology Division, Oak Ridge National Laboratory, Oak Ridge, TN, 37831, USA}
\email{lawriebj@ornl.gov}

%\keywords{hBN, spin defects, cathodoluminescence}

\footnotetext{This manuscript has been authored by UT-Battelle, LLC, under contract DE-AC05-00OR22725 with the US Department of Energy (DOE). The US government retains and the publisher, by accepting the article for publication, acknowledges that the US government retains a nonexclusive, paid-up, irrevocable, worldwide license to publish or reproduce the published form of this manuscript, or allow others to do so, for US government purposes. DOE will provide public access to these results of federally sponsored research in accordance with the DOE Public Access Plan (http://energy.gov/downloads/doe-public-access-plan).}

\begin{abstract}
The impulsive excitation of ensembles of excitons or color centers by a high-energy electron beam results in the observation of photon bunching in the second-order correlation function of the cathodoluminescence generated by those emitters. Photon bunching in cathodoluminescence microscopy can be used to resolve the excited-state dynamics and the excitation and emission efficiency of nanoscale materials, and it can be used to probe interactions between emitters and nanophotonic cavities. Here, we report substantial changes in the measured bunching induced by indirect electron interactions (with indirect electron excitation inducing $g^{2}(0)$ values approaching $10^4$).  This result is critical to the interpretation of $g^{2}(\tau)$ in cathodoluminescence microscopies, and, more importantly, it provides a foundation for the nanoscale characterization of optical properties in beam-sensitive materials.
\end{abstract}

\maketitle

\section{Introduction}

Cathodoluminescence (CL) microscopies have become workhorse tools for the characterization of nanomaterials with spatial resolution well beyond the optical diffraction limit \cite{abajoReview,Hayee2020,DOS_CL,garfinkel_2022,Sannomiya_nanocube,hachtel2019spatially}. As a result of near-field, high-energy, electron interactions, CL provides access to excitation pathways not accessible in conventional optical microscopies \cite{iyer2021near}.  Furthermore, time-resolved CL has emerged as a crucial tool for probing ultrafast dynamics of excited states in nanoscale materials \cite{TRCL_UF_TEM, CLExitation_pathways_2022}. Time resolved CL can rely on pulsed electron beams generated with electrostatic beam-blankers\cite{BB_CL} or pulsed-laser excitation of the electron gun \cite{Merano2005}, but both of these approaches reduce the spatial resolution of the microscope and add to the complexity of the microscope column design. Ultrafast beam blankers in scanning electron microscopes (SEMs) limit the temporal resolution to $\sim$ 100 picoseconds and provide spatial resolution of $\sim$ 50 nm \cite{Moerland:16_80psBB}. Pulsed laser excitation of the emitter leads to sub-ps electron pulses at the expense of poor spatial resolution\cite{MagdaThesis,Zewail_SUEM}.

Measures of the CL second order correlation function $g^{2}(\tau)$ in electron microscopes with continuous wave electron beams can provide access to nanoscale dynamics without recourse to pulsed electron sources\cite{MEURET201928,PhysRevLett.114.197401, garcia_Pulsed_statistics, PhysRevB.97.081404}. Moreover, spatially resolved measurements of $g^{2}(\tau)$ can be used to extract the excitation and emission efficiency with few-nanometer resolution \cite{g2mapping}.  
  
\begin{figure*}[hbt!]
    \centering
    \includegraphics[width=1.9\columnwidth]{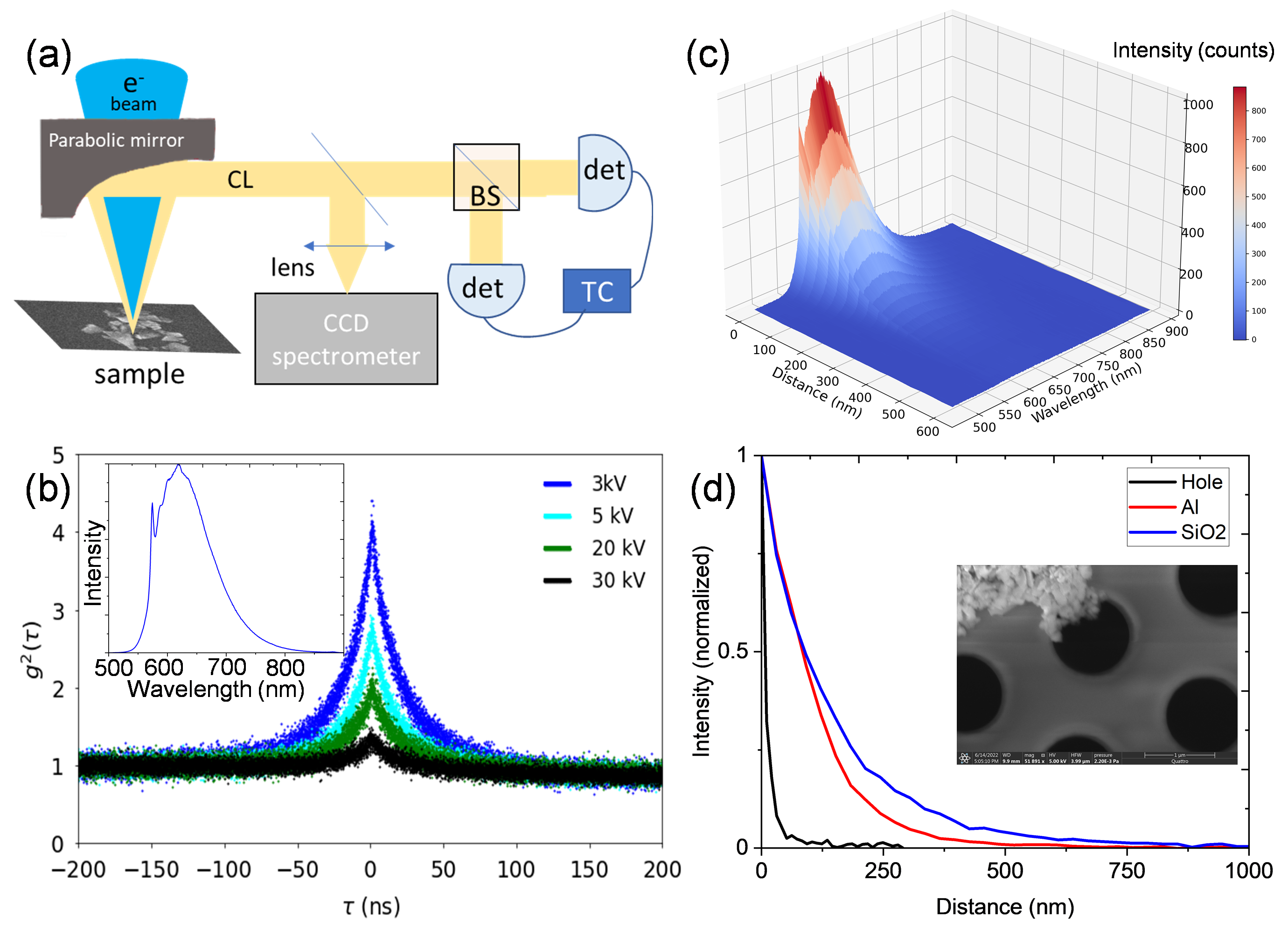}
    \caption{(a) Schematic of CL collection in an SEM, where CL spectrum imaging is performed with a CCD spectrometer and g\textsuperscript{2}($\tau$) measurement is performed using a beam splitter and two SNSPDs. A time correlator is used to time tag each detected photon. (b) g\textsuperscript{2}($\tau$) traces taken with electron beam parked on a nanodiamond crystal containing an ensemble of NV centers, showing decreasing bunching with increasing electron beam energy and 4.8 pA current. The inset shows the typical CL spectrum, with the peak at 575 nm being the NV$^0$ zero-phonon line. (c) Hyperspectral CL as a function of distance from a nanodiamond on an aluminum film at 5kV and 7 pA. (d) Intensity slice at 625 nm from CL spectra of nanodiamonds on Al and SiO\textsubscript{2} and nanodiamonds suspended on a TEM membrane as a function of distance from the nanodiamond (all at 5kV and 7pA). The inset shows a nanodiamond cluster at the edge of a hole in a SiN membrane.}
    \label{fig:Fig1}
\end{figure*}

CL bunching is typically understood as a result of high energy electron excitation of multiple plasmons, each of which can excite multiple defect centers within a sub-picosecond window \cite{CLExitation_pathways_2022}. In diamond, for example, high energy electrons can efficiently excite a 30-eV bulk plasmon mode \cite{diamond30evplasmon}. However, the indirect excitation mechanisms involved in the generation of bunched photons have not been investigated in detail so far. While bulk-plasmon mediated CL bunching has been explored carefully, the effect of plasmon mediated bunching for electron-beams incident on a substrate at some distance from color centers or excitons of interest has not been explored carefully. Similarly, the effect of aloof excitations and secondary electron excitations on CL bunching has not been described until now.

In this article, we study the effect of plasmons and secondary and aloof electron excitations on the measured CL bunching for nitrogen vacancy (NV) centers in nanodiamonds using g\textsuperscript{2}($\tau$) CL mapping supported by Monte-Carlo simulations and analytical modeling. We show that indirect excitation pathways can substantially modify the measured photon bunching, and we observe an exponential increase in g\textsuperscript{2}(0) as the electron beam moves away from the nanodiamond, reaching values as high as g\textsuperscript{2}(0)$\sim$1000, hundreds of times greater than the g\textsuperscript{2}(0) measured for the same electron beam directly incident on the nanodiamond. We study the role of the substrate on the measured photon bunching, and compare g\textsuperscript{2}($\tau$) for nanodiamonds deposited on metallic (aluminum) and insulating (SiO\textsubscript{2}) films. We also probed aloof excitation pathways directly with nanodiamonds suspended over holes in a SiN membrane. 

\section{Experimental Methods}

%The electron beam in our scanning electron microscope (SEM) has a spot size of 0.7 nm. The CL spatial resolution is dictated by the beam spot size and the free-carrier migration after electron-beam excitation. Even so, CL offers 1-2 orders of magnitude higher spatial resolution than conventional PL microscopy. 

Fluorescent nanodiamonds (Adamas Nano) with an average particle size of 140 nm were dropcast on a SiN membrane and on a patterned sample comprising 50 nm thick aluminum pads on a continuous 300 nm thick SiO\textsubscript{2} film on a silicon substrate (which allowed for direct comparison of the NV CL photon statistics on Al and on SiO\textsubscript{2} under identical imaging conditions). Each nanoparticle contains $\sim$ 1200 NV\textsuperscript{0} centers; this defect density is high enough to suppress any spatial heterogeneity in the CL emitted from the nanodiamond, consistent with other reports in the literature \cite{TEM_NV_CL_Kociak}. 

CL microscopy was performed in a FEI Quattro SEM with a Delmic Sparc CL collection system. The electron beam excites the sample through a hole in a parabolic mirror. The mirror collimates the CL emission, which is sent to an Andor Kymera spectrometer for spectrum mapping or to a Hanbury Brown-Twiss Interferometer for g\textsuperscript{2}($\tau$) measurements. The Hanbury Brown-Twiss Interferometer comprises a fiber beamsplitter and a pair of Quantum Opus large area superconducting nanowire single photon detectors (SNSPDs), as shown in Fig. \ref{fig:Fig1}(a). Detected photons are time tagged by a HydraHarp 400 time-interval analyzer with 128 ps bin sizes. The secondary electron (SE) signal is simultaneously measured on an Everhart-Thornley detector. 

%The second order correlation function g\textsuperscript{2}($\tau$), which measures the correlation between photons detected at time t and $t+\tau$, is defined as, 
%\begin{equation}
%g\textsuperscript{2}(\tau) = \frac{\langle I(t)I(t+\tau)\rangle}{\langle I(t)\rangle \langle I(t+\tau)\rangle}     
%\end{equation}

\section{Results}

The inset in Fig. \ref{fig:Fig1}(b) shows a typical CL spectrum from the nanodiamond, with the distinctive NV\textsuperscript{0} peak at 575 nm and a broad phonon sideband at longer wavelengths. The NV\textsuperscript{-} peak is suppressed under electron-beam excitation due to beam-induced charge-state conversion \cite{ultrafastCL}. The change in measured bunching as a function of electron-beam energy is shown in Fig. \ref{fig:Fig1}(b) for a single nanodiamond crystal on the aluminum film. The bunching decreases with increasing energy at constant 7 pA current, as previously observed in literature. \cite{PhysRevLett.114.197401}.

We next examine the CL response of a nanodiamond on an aluminum film when the electron beam is not directly incident on the nanodiamond as shown in Fig. \ref{fig:Fig1}(c) (where zero distance is defined as the edge of the nanodiamond). The CL intensity decays uniformly over a distance of several hundred nanometers, consistent with a combination of well-understood indirect excitation pathways, including secondary electron excitation, aloof excitation and plasmon-mediated excitation. The normalized CL intensity at a wavelength of 625 nm is plotted as a function of distance for nanodiamonds on Al and SiO\textsubscript{2} films and a holey SiN membrane in Fig. \ref{fig:Fig1} (d). The inset shows suspended nanodiamonds on a holey SiN membrane. The length scales for aloof excitations over a SiN hole are substantially smaller than those for plasmon- and secondary-electron mediated excitation, but in each case, substantial CL intensity can be measured with indirect electron-beam excitation.

\begin{figure}[ht!]
    \centering
    \includegraphics[width=\columnwidth]{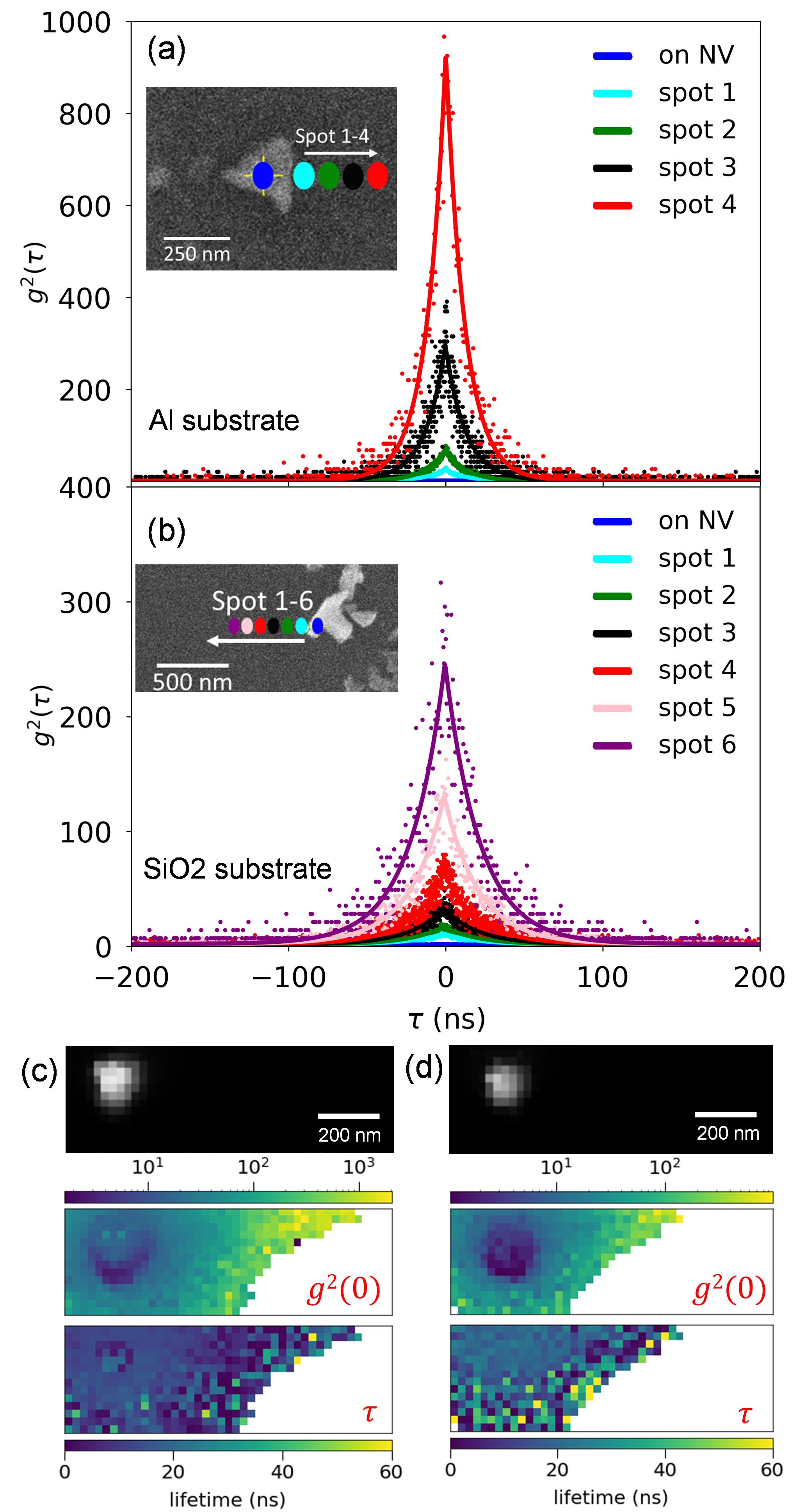}
    \caption{Measured NV CL g\textsuperscript{2}($\tau$) for various positions on (a,c) Al and (b,d) SiO\textsubscript{2} films. The magnitude of g\textsuperscript{2}(0) dramatically increases as the beam moves away from the nanodiamond in both cases. (c) and (d) show the 2D bunching map collected with 10 sec integration/pixel and a 25 nm pixel size, including the panchromatic CL intensity (top panels), the measured CL bunching (middle panels) and the measured bunching time constant (bottom panels). All data was acquired with a 5kV, 7pA electron beam.}
    \label{fig:Fig2}
\end{figure}

We investigate the spatially resolved bunching dynamics in a similar fashion, i.e. as a function of electron-beam distance from the nanodiamond. The second order correlation functions measured over a 120 second integration time at 4(6) spots for nanodiamonds on Al (SiO\textsubscript{2}) are shown in Fig. \ref{fig:Fig2}a(b). In both cases, we observe a dramatic increase in g\textsuperscript{2}(0) with increasing distance from the nanodiamond. The NV center lifetime, obtained by fitting g\textsuperscript{2}($\tau$) curve \cite{MEURET201928} with a single exponential decay, exhibited no statistically significant dependence on position, with an average value of 25 $\pm$ 1 ns on Al and 43 $\pm$ 1 ns on SiO\textsubscript{2}. The lifetimes are similar to those observed previously in literature\cite{Liaugaudas_NV_lifetime,Storteboom_NV_lifetime} and the faster lifetime on the aluminum substrate can be attributed to either a plasmon-mediated Purcell enhancement\cite{dielectric_APL_NV,Purcell_CL} or to hot-electron transfer between the diamond and the metallic film\cite{wieghold2020probing,lawrie2009enhancement}. We also collected 2D bunching maps with 25 nm pixel resolution and 10 sec/pixel integration times. As shown in Fig. \ref{fig:Fig2}(c,d), g\textsuperscript{2}(0) increases exponentially with increasing distance from the nanodiamond (middle panel) and little change was observed in the time constant (lower panel). The nanodiamonds are located at the bright spots in the panchromatic CL images shown in the top panel of Fig. \ref{fig:Fig2}(c,d) .  

\begin{figure}[ht!]
    \centering
    \includegraphics[width=\columnwidth]{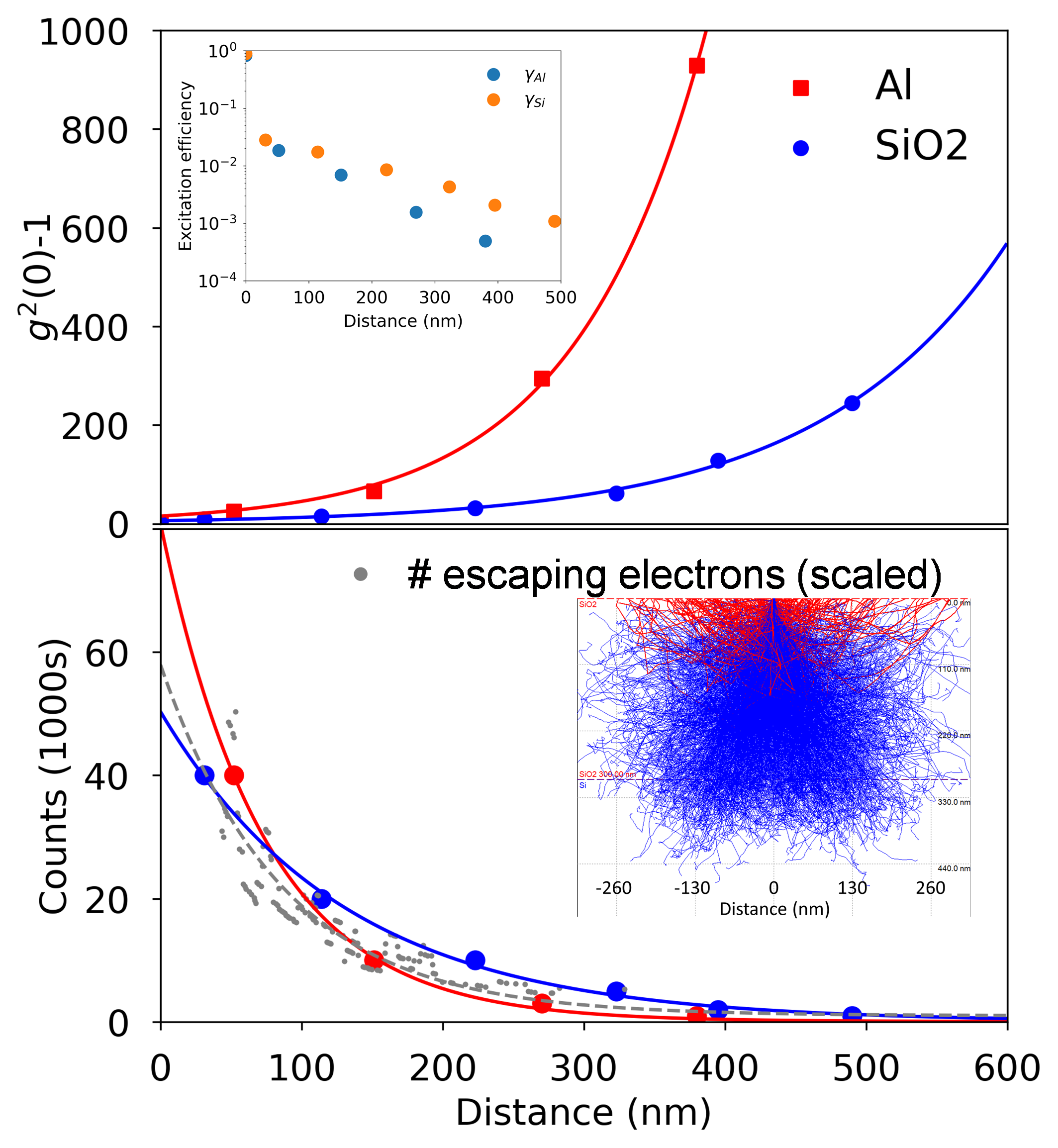}
    \caption{(top panel) g\textsuperscript{2}(0) as a function of distance from nanodiamonds on Al and SiO\textsubscript{2} substrates . The inset shows the excitation efficiency, $\gamma$ calculated using equation \ref{eq:2} and \ref{eq:3}. (bottom panel) Single channel counts/sec. The solid lines are exponential fits. (Inset of bottom panel) Electron trajectories from Casino Monte-Carlo simulation, where red trajectories illustrate the escaping secondary electrons. The number of escaping electrons (scaled) as a function of distance is shown in grey. All data was acquired with a 5kV, 7pA electron beam.}.
    \label{fig:g2-counts}
\end{figure}

The top panel of Fig. \ref{fig:g2-counts} illustrates the position dependence of NV CL bunching extracted from exponential fits to the g\textsuperscript{2}($\tau$) shown in Fig.~\ref{fig:Fig2}(a,b). A least squares fit to the extracted bunching illustrates the exponential growth in the measured bunching as a function of distance from the nanodiamond. On the other hand, the photon counts recorded by the SNSPDs exhibit an exponential decay as shown in the bottom panel of Fig.\ref{fig:g2-counts}. The changes in g\textsuperscript{2}(0) can be analytically modeled to assign an effective excitation efficiency as a function of electron beam distance from the nanodiamond. An analytical model developed by  Sol\`{a}-Garcia et al. has been previously used to describe g\textsuperscript{2}($\tau$) as a function of current, lifetime and excitation efficiency \cite{garcia_Pulsed_statistics}.  We use that framework here with the assumption that g\textsuperscript{2}($\tau$) can be fit to a single exponential decay, and we model the second order correlation function as:

\begin{equation} \label{eq:2}
g\textsuperscript{2}(0) = 1 + \frac{q}{2I*\tau_{emitter}}\left( \frac{b+1}{b} \right)   
\end{equation}
where q is the electron charge, I is the current, $\tau_{emitter}$ is the emitter lifetime, and b is the expected number of bulk plasmons excited per electron. The excitation efficiency, $\gamma$ can be defined as\cite{garcia_Pulsed_statistics}:

\begin{equation} \label{eq:3}
\gamma = 1 - e^{-b}.   
\end{equation}

\begin{figure}[ht!]
    \centering
    \includegraphics[width=\columnwidth]{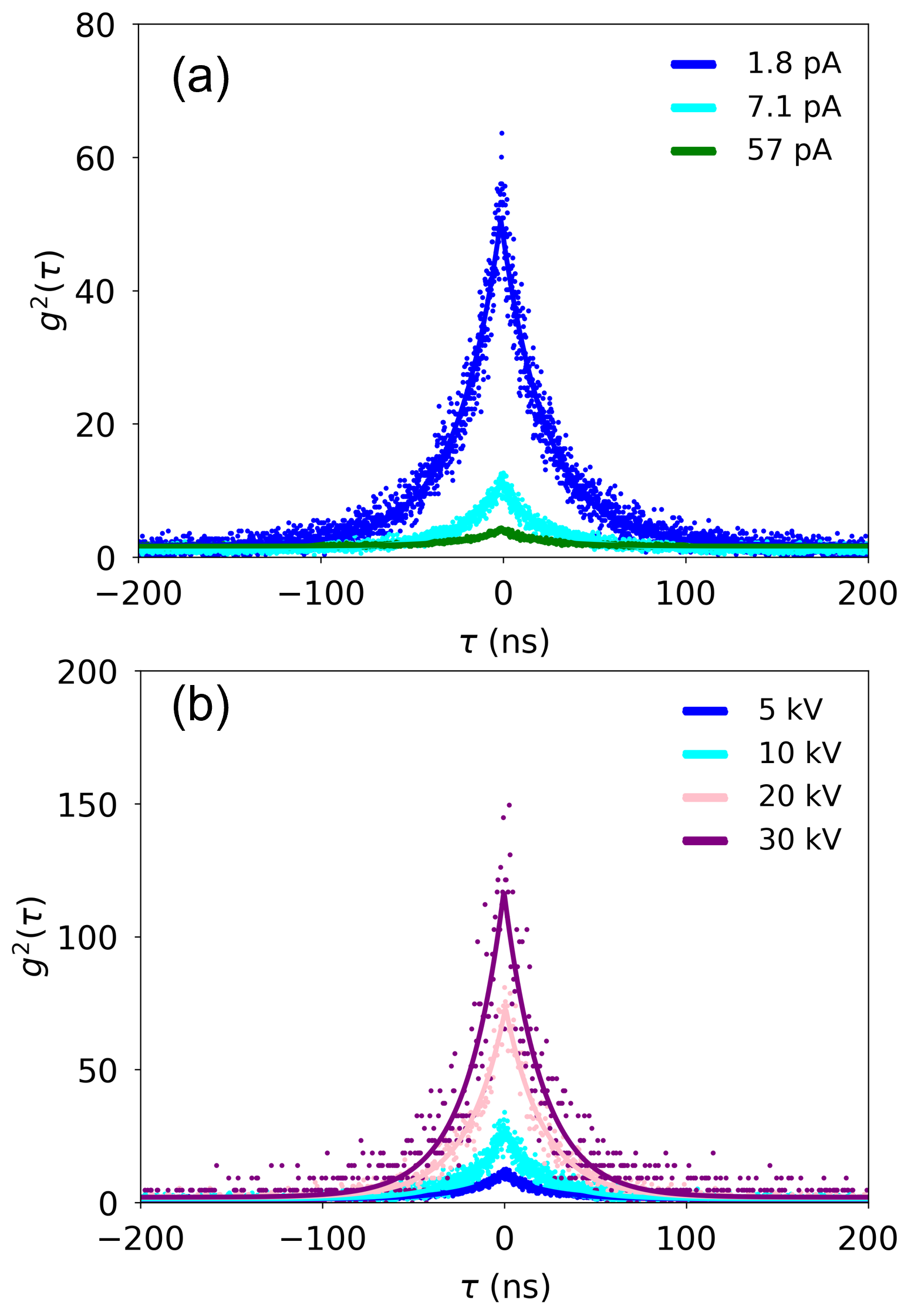}
    \caption{(a) Current dependence at 5 kV accelerating voltage and (b) accelerating voltage dependence at $\sim$10 pA of g\textsuperscript{2}($\tau$) for electron beam parked at 30 nm distance from the nanodiamond. }
    \label{fig:I_and_V_dep}
\end{figure}

The position dependent excitation efficiency is plotted in the inset of Fig.\ref{fig:g2-counts} (top panel), demonstrating a clear exponential decay of excitation efficiency with distance. To obtain further insight into the indirect excitation mechanisms, a Monte-Carlo simulation is performed in Casino to model the density of secondary electrons as a function of distance from the point of excitation\cite{drouin1997casino}. The inset in the bottom panel of Fig.\ref{fig:g2-counts} depicts the electron trajectories in the simulation. The red trajectories are the electrons that can escape the substrate and excite the nanodiamond. The extracted number of electrons are plotted in grey and scaled to match the counts at 40 nm distance. The position dependence of the number of secondary electrons that interact with the nanodiamond is qualitatively consistent with the position dependence of the NV CL counts. We note that the distance dependence of secondary electrons interacting with the nanodiamond was very similar for both Al and SiO\textsubscript{2} substrates, and hence we have only shown the simulation for SiO\textsubscript{2}.

Finally, we investigated the current and voltage dependence of g\textsuperscript{2}($\tau$) for an electron beam parked at a fixed distance of 30 nm from the nanodiamond on the SiO\textsubscript{2} substrate. We found increasing g\textsuperscript{2}(0) with decreasing current (Fig. \ref{fig:I_and_V_dep}(a)), as observed previously in literature. However, we also observed increasing g\textsuperscript{2}(0) with increasing voltage  (Fig. \ref{fig:I_and_V_dep}(b)), in marked contrast to previous reports in the literature and to measurements reported here with the electron-beam directly incident on the nanodiamond, as shown in Fig.\ref{fig:Fig1}(b)~\cite{PhysRevLett.114.197401}. 

This anomalous scaling can be explained as a result of the energy dependence of the excitation efficiency in the aloof-excitation regime. When a nanodiamond is directly irradiated by high energy electrons, its bulk plasmon mode exhibits a large cross section for excitation through momentum-transferring inelastic scattering processes. In contrast, optical selection rules take effect for aloof excitations, precluding any bulk plasmon mode excitations when the beam is focused well outside of the nanodiamond.  Only optically allowed transitions will be promoted in this case, with the probability of inducing a given excitation (within the electric-dipole approximation \cite{schatz2002quantum}) proportional to the squared modulus of the inner product between its transition dipole moment and the electric field emanated by the distant passing electron.  The observed increase in g\textsuperscript{2}(0) for CL from an ensemble of quasi-impulsively excited color centers with increasing beam energy is therefore consistent with the electric field emanated by the beam becoming stronger in the directions transverse to its motion for faster electrons (a consequence of relativistic length contraction that is described for a charged particle undergoing uniform motion within the Li\'enard-Wiechert formalism\cite{feynman1965feynman}).

\section{Conclusion}
In conclusion, a clear understanding of indirect electron-beam excitation processes is critical to the understanding of CL g\textsuperscript{2}($\tau$) microscopies. We found that the measured bunching increases exponentially as the electron beam is moved away from an ensemble of emitters, with a corresponding exponential decrease in the total number of counts measured. The distance dependence of the bunching suggests that the reduced excitation efficiency for color centers excited by secondary electrons is the primary cause of the unexpectedly high photon bunching, though aloof interactions introduce a similar effect at shorter length scales. Cathodoluminescence microscopies have become increasingly critical tools for probing beam-sensitive 2D materials\cite{nayak2019cathodoluminescence,curie2022correlative} and perovskite films\cite{xiao2015mechanisms,taylor2022hyperspectral}.  The ability to probe nanoscale excited state dynamics in these materials without direct electron-beam excitation will enable new CL imaging modalities that would otherwise introduce unmanageable beam-induced damage due to excessive dwell times.

\section{Acknowledgments}
This research was supported by the Center for Nanophase Materials Sciences, which is a U.S. Department of Energy Office of Science User Facility. J.N. was supported by the U.S. Department of Energy, Office of Science, Office of Workforce Development for Teachers and Scientists (WDTS) under the Science Undergraduate Laboratory Internship program.

%\bibliography{references}
%apsrev4-2.bst 2019-01-14 (MD) hand-edited version of apsrev4-1.bst
%Control: key (0)
%Control: author (8) initials jnrlst
%Control: editor formatted (1) identically to author
%Control: production of article title (0) allowed
%Control: page (0) single
%Control: year (1) truncated
%Control: production of eprint (0) enabled
%

\end{document}